\documentclass[aps,showpacs,a4paper,floatfix,twocolumn,prc,amsmath,amssymb]{revtex4-1}
\usepackage{amsmath}
\usepackage{graphicx}
\usepackage{ulem}
\usepackage{morefloats}
\usepackage{rotating}
\usepackage{relsize}
\usepackage{color}
\usepackage{floatflt,epsfig}
\usepackage{ulem}
\usepackage{natbib}
\usepackage{hyperref}
\usepackage{filecontents}
\begin{document}

\title{Combined Rastall and Rainbow theories of gravity with applications to neutron stars}

\author{Cl\'esio E. Mota $^1$}
\email{clesio200915@hotmail.com}
\author{Luis C. N. Santos$^2$}
\email{luis.santos@ufsc.br}
\author{Guilherme Grams$^1$}
\email{grams.guilherme@gmail.com}
\author{Franciele M. da Silva$^1$}
\email{franmdasilva@gmail.com}
\author{D\'ebora P. Menezes $^1$}
\email{debora.p.m@ufsc.br}

\affiliation{$^1$Departamento de F\'isica, CFM - Universidade Federal de Santa Catarina; C.P. 476, CEP 88.040-900, Florian\'opolis, SC, Brasil}
\affiliation{$^2$Departamento de F\'isica, CCEN - Universidade Federal da \\ Para\'iba; C.P. 5008, CEP  58.051-970, Jo\~ao Pessoa, PB, Brasil.}

\begin{abstract}
The possibility of modifications on general relativity is investigated. We propose
an alternative theory of gravity constructed with the combination of Rastall
and Rainbow theories. The hydrostatic equilibrium equations are obtained
in order to test the new theory in neutron stars, whose mass-radius diagrams are obtained 
using modern equations of state of nuclear matter derived from relativistic mean field models and compared with the ones computed by the Tolman-Oppenheimer-Volkoff equations. We conclude that substantial modifications are obtained even for very small alterations on the two free parameters, making the reproduction of astrophysical observations an easy task.
\\
\\
\noindent Keywords : general relativity, modified gravity, neutron stars. 
\end{abstract}

\maketitle

\section{Introduction}
\label{sec:intro}

The study of compact objects is an interdisciplinary subject that requires the understanding of many topics and two essential areas are gravitational theory 
and nuclear physics. Neutron stars equilibrium conditions are guaranteed by the balance between the nuclear degeneracy pressure and the strong gravitational field, produced by its very dense matter that works in order to hold stellar matter together.

The theory of gravity proposed by Einstein one century ago has helped us to comprehend many aspects of the universe and it keeps passing in every test \cite{RGtests} up to today. Notice the recent observation of
gravitational waves by binary black hole \cite{GWblackhole}
and neutrons star \cite{GW170817} mergers and the first photography of a black hole obtained by the Event Horizon Telescope \cite{10April}.
Despite the success of general relativity (GR), alternative theories of gravity
have been proposed in the last decades \cite{alternativegravity}. Some arguments for these theories come from the assumption that
the equivalence principle (necessary to Einstein GR) could cease to be valid at large distances. Moreover, it has been hypothesized that 
the rotation velocity of our galaxy could be explained without dark matter and
the accelerated expansion of the Universe could be obtained without dark energy if we upgraded our theory of gravity beyond general relativity \cite{Olmo07,Hui09,Bertolami07,Tarko10,Abdalla05,Cognola08}.

In 1972 Peter Rastall proposed a generalization of Einstein GR \cite{rastall72}.
Rastall questioned the validity of the conservation law of the energy-momentum tensor in curved space time. In this new theory of gravity the covariant derivative of $T_{\mu \nu}$ does not vanish
and depends on the curvature $R$ and on a free parameter. We show in Sec. \ref{Rastalltheory} the main ideas underlying the modifications on Einstein theory proposed by Rastall.

In \cite{fabris_rastall} the authors investigated the effects of Rastall theory on neutron stars modeled with both polytropic and  non relativist equations of state (EoS). They concluded that only small deviations of GR are consistent with neutron star constraints and only values of the free parameter $\lambda$ bigger then one were tested due to considerations on the energy conditions. We will go back to this point when testing our model.

In 2004, another theory called gravity rainbow, which is an extension of the doubly (or nonlinear or deformed) special relativity for curved space-times
was proposed by Jo\~ao Magueijo and Lee Smolin
\cite{magueijo}. The principles of the doubly special relativity \cite{bruno,kowalski,giovanni,smolin,magueijo1} are: (i) the laws of physics are the same in all inertial frames; (ii) in the limit $\frac{E}{E_{p}} \rightarrow 0$, where $E_{p}=\frac{\sqrt{\hbar c^{5}}}{G}$ is the Planck energy, the speed of massless quanta goes to $c$, for all inertial frames; (iii) $E_{p}$ is an universal constant for all inertial frames. By making this extension they concluded that the geometry of the space-time becomes energy dependent, so that, families of one-parameter metrics parametrized by $\frac{E}{E_{p}}$ are possible, where each family can be referred as a rainbow metric. In this new theory, the authors obtained a cosmological solution that leads to an energy dependent age for the universe, which could solve the horizon problem. They also obtained a solution analogue to Schwarzschild solution, in which the area of the event horizon is energy dependent, with important consequences on the black-hole thermodynamics. 

In \cite{garattini,hendi} the authors obtained TOV-like solutions to this theory and one of their conclusions was that in gravity rainbow neutron stars with maximum masses larger than $2 M_{\odot}$ are easily attainable, as expected from observational results \cite{Demorest,Antoniadis}. 

The proposal of this work is to merge Rastall and Rainbow theory in a new post-GR theory, which we call from now one  Rastall-Rainbow (RR) theory. We consider the Einstein's field equations modified by Rastall and then change these field equations so that the space-time metric becomes dependent on the test particles energy, in accordance with Rainbow gravity. Thus, all the quantities that compose the Rastall field equations become dependent on the particles energy. 

As a first test to Rastall-Rainbow theory we derive a new set of equations that describe the star equilibrium and  investigate the mass-radius relation of neutron stars. We obtain the
stellar properties of a family of neutron stars using as input modern relativistic equations of state. 

From the microphysics point of view, the nuclear physics community has made a big effort in the past years to construct appropriate equations of state (EoS) to describe stellar matter which generates the nuclear pressure that holds the neutron stars from gravitational collapse.
For this task, one can work with different formalisms, and the two most common ways to obtain EoS are the non relativistic \cite{Dutra2012}, which usually use a modified liquid drop model with Skyrme interaction to simulate the nucleon interaction,
and the relativistic nuclear models \cite{walecka,bogutabodmer,guichon}, 
which are originated from mean-field theory (MFT) applied to Lagrangian densities. 
For the present work, we choose 3 different EoS constructed with relativistic models. The microphysics of our {\it 'gravity lab'} will be modeled with one EoS derived from the Walecka model corrected to include non-linear terms  \cite{walecka,bogutabodmer}, the IU-FSU model \cite{iufsu} and two EoS derived from the quark-meson coupling (QMC) model  \cite{guichon}, a relativistic nuclear model that takes into account quark degrees freedom.

We use the above mentioned EoS to compute the macroscopic properties of neutron stars by testing a new alternative theory of gravity (RR) because they 
have already been confronted with nuclear matter bulk properties and stellar constraints
\cite{Debi12,Orsaria14,Fattoyev18,Miyatsu13,Panda04,Panda12,Grams18, Dutra2014,Dutra2016}. 

The paper is organized as follows: In Sec. \ref{RRtheory} we make a brief review of the
original Rastall and Rainbow theories and present the formalism 
that allows us to construct a new theory that results from the combination
of the two alternative gravity theories. Section \ref{TOV_RR} is reserved to deduce the new equilibrium conditions
for neutron stars. We show our results and discussions in Sec. \ref{results} and
draw the final conclusions in Sec. \ref{conclusion}.

\section{Formalism}
\label{RRtheory}

In this section we discuss the unification of Rastall and Rainbow theories. The idea is to generalize the Rastall gravity to an energy dependent Rastall theory. Of course, the final form of the field equation is an expression that captures elements of both theories. It is expected that the resulting theory will be able to reproduce the key features of Rainbow gravity as the modification of the dispersion relation near the Planck scale besides explaining astronomical and cosmological phenomena such as the threshold anomalies of ultra high energy cosmic rays. In addition, the field equations must incorporate the change in the conservation law of the energy-momentum tensor according to the Rastall theory. As a test for the theory, we solve the field equation originated from the unification of Rastall and Rainbow theories to study mass and radii of neutron stars.

\subsection{Rainbow gravity}
\label{Rainbowtheory}

Doubly special relativity theory with an invariant energy scale may be generalized to general
relativity curved space-time. Initially, we consider a deformation of the usual dispersion relation
\begin{equation}
    E^{2}\Xi(x)^{2}-p^{2}\Sigma(x)^{2}=m^{2},
\label{eq1}
\end{equation}
where the argument $x=E/E_{p}$ is the ratio of the energy of a probe particle to the Planck energy $E_{p}$. In Eq. (\ref{eq1}) the functions $\Xi$ and $\Sigma$ are called rainbow functions. In the literature, the choice of the these functions is theoretically and phenomenologically motivated. For example, in \cite{khodadi,Adel} the authors obtain solutions corresponding to a nonsingular universe through the choice of functions $\Xi=1$ and $\Sigma=\sqrt{1+x^{2}}$. On the other hand, an exponential form of rainbow is applied in the study of gamma ray burst \cite{Adel,Amelino}. It is important to note that in the infrared limit the standard energy--momentum dispersion relation is recovered in this type of theory. In this way, the functions $\Xi$ and $\Sigma$  satisfy the
conditions:
\begin{equation}
    \lim_{x\rightarrow 0} \Xi (x)=1, \quad \lim_{x \rightarrow 0} \Sigma (x)=1.
    \label{eq2}
\end{equation}
In the absence of gravity, the space-time acquires a geometry that depends on the energy of the particles. In the presence of gravity, the space-time metric can be constructed using the following energy dependent metric \cite{magueijo}
\begin{equation}
    g(x)=\eta^{ab} e_{a}(x)\otimes e_{b}(x),
    \label{eq3}
\end{equation}
where the energy dependent frame fields $e_{a}(x)$ are related to the energy independent frame fields, denoted by $\widetilde{e_{a}} (x)$, as follows:
\begin{equation}
    e_{0}(x)=\frac{1}{\Xi (x)} \widetilde{e_{0}}, \quad e_{i}(x)=\frac{1}{\Sigma (x)} \widetilde{e_{i}}.
    \label{eq4}
\end{equation}
In Rainbow gravity, the Einstein equations are replaced by one parameter family of field equations due to the modification of the dispersion relation. Therefore, by using Eq. (\ref{eq3}) and considering the usual general relativity quantities  $\widetilde{e_{i}}$,  we obtain an energy dependent metric with spherical symmetry in the form
\begin{equation}
    ds^{2}=-\frac{B(r)}{\Xi^{2}} dt^{2}+\frac{A(r)}{\Sigma^{2}} dr^{2}+\frac{r^{2}}{\Sigma^{2}}(d\theta^{2}+\sin{\theta}^{2}d\phi^{2}),
    \label{eq5}
\end{equation}
where $A(r)$ and $B(r)$ are radial functions. Thus, the spherically symmetric metric depends on the energy due to the rainbow functions. Note that in this configuration the coordinates $r$, $t$, $\theta$ and $\phi$ are independent of the energy of the probe particles. The next step is to study the effect of the energy dependence in the context of Rastall gravity. For this purpose we introduce the main ideas of this theory in the following pages.

\subsection{Rastall gravity}
\label{Rastalltheory}
It is true that the left side of the usual Einstein's field equations satisfies $G^{\mu \nu}_{\quad;\mu}=0$, which can be easily verified by using the Bianchi identities. This relation is in accordance with the right side of the field equation if $T^{\mu \nu}_{\quad;\mu}=0$. However, there is another way to write the covariant derivative of the energy-moment tensor keeping both sides of the Einstein's equation coherent with each other. Peter Rastall proposed a modification of the conservation law of the energy-momentum tensor in curved space-time in the form \cite{rastall72}:

\begin{equation}
    T_{\: \: \mu;\nu}^{\nu}=\Bar{\lambda}R_{,\mu},
    \label{eq6}
\end{equation}
where $\Bar{\lambda}$ is an undetermined constant. From Eq. (\ref{eq6}), we can write 
\begin{equation}
    \left(T_{\: \: \mu}^{\nu}-\Bar{\lambda}\delta_{\: \:\mu}^{\nu}R\right)_{;\: \nu} =0.
    \label{eq7}
\end{equation}
In fact, the assumption (\ref{eq6}) is consistent with the field equations 
\begin{equation}
  R^{\nu}_{\: \: \mu}-\frac{1}{2}\delta_{\: \:\mu}^{\nu}R= 8\pi G\left(T_{\: \: \mu}^{\nu}-\Bar{\lambda}\delta_{\: \:\mu}^{\nu}R\right),  
  \label{eq8}
\end{equation}
a modified Einstein's field equation. It is useful to rewrite this equation so that only the energy-moment tensor stays on the right side, i.e.,
\begin{equation}
    R^{\nu}_{\: \: \mu}-\frac{\lambda}{2}\delta_{\: \:\mu}^{\nu}R= 8\pi G T_{\: \: \mu}^{\nu},
  \label{eq9} 
\end{equation}
where we have defined $\Bar{\lambda}=\frac{1-\lambda}{16\pi G}$. When $\lambda =1$, the usual field equation is reobtained. Thus, the parameter $\lambda$ is related to the generalization of the Einstein's equation. In the flat space-time, when $R = 0$, Eq. (\ref{eq6}) recovers the usual conservation law. The change proposed by Rastall has effects in the case of  general space-time.

\subsection{Rastall-Rainbow theory}
\label{TOV_RR}

Above we have discussed the modified gravitational theories separately. It is interesting to study both theories in a unified formalism. For this purpose, the starting point is the modified Einstein's equation given in Eq. (\ref{eq9}) and the modified conservation law (\ref{eq6}). The effect of Rainbow gravity can be incorporated into Eq. (\ref{eq9}) by considering an energy dependent metric and an energy dependent gravitational constant $G(x)$, resulting in the Rastall-Rainbow field equations in the form
\begin{equation}
  R^{\nu}_{\: \: \mu}(x)-\frac{\lambda}{2}\delta_{\: \:\mu}^{\nu}R(x)= 8\pi G (x)T_{\: \: \mu}^{\nu}(x)  
  \label{eq10},
\end{equation}
where the Rastall parameter $\lambda$ is energy independent. We may solve Eq. (\ref{eq10}) in the energy dependent metric with spherical symmetry defined in Eq. (\ref{eq5}). This space-time can be used to model the internal structure of a star and we next obtain a new set of equations that describe stellar equilibrium, i.e., the modification of the  Tolman--Oppenheimer--Volkoff (TOV) equations \cite{TOV1,TOV2} due to the Rastall-Rainbow gravity. 
When we consider the usual General Relativity (GR), the solution of Einstein's field equation allows us to understand the hydrostatic equilibrium of homogeneous, static, isotropic and spherically symmetric objects. In particular, we are interested in compact objects such as neutron stars. It is expected that the influence of the energy dependence of the metric and of the change in the conservation law of the energy-moment tensor, will modify the usual relations of the hydrostatic equilibrium inside these compact objects. As discussed above, we consider a line element with spherical symmetry as
given in Eq. (\ref{eq5}). 
We assume that matter in the stellar interior can be described by the tensor energy-moment of a perfect fluid in a co-moving frame, usually written as
\begin{equation}
    T_{\mu\nu}=pg_{\mu\nu}+(p+\rho)U_{\mu}U_{\nu}
    \label{eq12},
\end{equation}
where $p(r)$ and $\rho(r)$ are respectively the pressure and the energy density of the fluid. The term $U_{\mu}$, that satisfies $U_{\mu}U^{\mu}=-1$, is the 4-velocity of the a fluid element defined as 
\begin{equation}
    U_{\mu}=\left( \frac{\Xi(x)}{\sqrt{B(r)}},0,0,0\right)
    \label{eq13}.
\end{equation}
 Adding and subtracting the term $(1/2) g_{\mu\nu}R$ to the left side of Eq. (\ref{eq10}), in its covariant form, it can be written as the usual Einstein equation with an effective energy-moment tensor, i.e.,
 \begin{equation}
     R_{\mu\nu}-\frac{1}{2}g_{\mu\nu}R=8\pi G\tau_{\mu \nu},
     \label{eq13b}
 \end{equation}
 where we have defined 
 \begin{equation}
     \tau_{\mu\nu}=T_{\mu\nu}-\frac{(1-\lambda)}{2(1-\lambda)}g_{\mu\nu}T.
      \label{eq13c}
 \end{equation}
Note that the trace of the energy-moment tensor $T=(1-2\lambda)R/8\pi G$ has been used to replace $R$ in the right side. Now the usual energy-moment tensor of a perfect fluid (\ref{eq12}) together with expression for the 4-velocity (\ref{eq13}) may be used on the right side of the field equation (\ref{eq13b}). The result reads:
\begin{equation}
 -\frac{B}{r^{2}A} + \frac{B}{r^{2}} + \frac{A'B}{rA^{2}} = 8\pi GB \Bar{\rho},  \label{eq14}
\end{equation}

\begin{equation}
-\frac{A}{r^{2}} + \frac{B'}{rB} + \frac{1}{r^2}  = 8\pi GA \Bar{p},  \label{eq15}
\end{equation}

{\fontsize{9.7}{12}
\begin{align}
 -\frac{B'^{2}r^{2}}{4AB^{2}} - \frac{A'B'r^{2}}{4A^{2}B} + \frac{B'' r^{2}}{2AB} - &\frac{A'r}{2A^{2}} + \frac{B'r}{2AB} \nonumber \\
& = 8\pi Gr^{2} \Bar{p},  \label{eq16}
\end{align}}
where $\Bar{\rho}$ and $\Bar{p}$ are effective pressure and energy density defined in the form
\begin{align}
    \Bar{\rho} & = \frac{1}{\Sigma(x)^{2}}\left[\alpha_{1}\rho+3\alpha_{2}p\right],\label{eq17}\\
    \Bar{p} & = \frac{1}{\Sigma(x)^{2}}\left[\alpha_{2}\rho+(1-3\alpha_{2})p\right], \label{eq18}
\end{align}
where
\begin{equation*}
    \alpha_{1}=\frac{1-3\lambda}{2(1-2\lambda)}; \qquad \alpha_{2}=\frac{1-\lambda}{2(1-2\lambda)}.
\end{equation*}
The set of coupled differential equations (\ref{eq14})$-$(\ref{eq16}) are similar to the ones obtained in the GR. In this way, it is possible to find the form of the function $A(r)$ through a direct integration of Eq. (\ref{eq14}), providing the expression
\begin{equation}
    A(r)=\left[1-\frac{2GM(r)}{r} \right]^{-1},
    \label{eq19}
\end{equation}
together with the definition of the mass term
\begin{equation}
    M(r)=\int_{0}^{R} 4\pi r'^{2}\Bar{\rho}(r')dr'.
    \label{eq20}
\end{equation}
The interpretation of Eq. (\ref{eq20}) is direct: The integral is performed from the stellar center to $r=R$. Therefore $R$ denotes the radius of the star, where $\Bar{\rho}$ is an effective energy density in the star interior. In the case where $\lambda = 1$ and $\Sigma=1$ we have $\Bar{\rho}=\rho$, then this definition coincides with the usual definition of the GR. Note that the gravitational mass $M(r)\equiv M_{G}$  defined in  (\ref{eq20}) is obtained using the effective density defined in Eq. (\ref{eq17}). In this way, the Rainbow function $\Sigma$ and Rastall parameter $\lambda$ modify the stellar mass in this formalism. At this stage, we make use of the modified  conservation law for the energy-momentum tensor $ T_{\: \: \mu;\nu}^{\nu}=\Bar{\lambda}R_{,\mu}$ to obtain
\begin{equation}
 \frac{B'}{2B}  = -\frac{\Bar{p}'}{\Bar{p}+\Bar{\rho}}. 
 \label{eq21}
\end{equation}
 Manipulating Eq. (\ref{eq15}), employing the result (\ref{eq19}), we derive the equation
\begin{align}
 \frac{B'}{2B} & = \frac{GM}{r^{2}} \left[1+\frac{4\pi r^{3} \Bar{p}}{M} \right]\left[1-\frac{2GM}{r}\right]^{-1}. \label{eq22} 
\end{align}
Finally, we can eliminate function $B$ by identifying Eq. (\ref{eq21}) with Eq. (\ref{eq22}) and then isolating the term $\Bar{p}'$. The result is
\begin{equation}
\Bar{p}'= -\frac{GM\Bar{\rho}}{r^{2}}\left[1+\frac{\Bar{p}}{\Bar{\rho}}\right] \left[1+\frac{4\pi r^{3} \Bar{p}}{M} \right]\left[1-\frac{2GM}{r}\right]^{-1}, 
\label{eq23}
\end{equation}
This equation gives us information about the stellar hydrostatic equilibrium within the context of Rastall-Rainbow gravity. The effective pressure and density are physical quantities that depend on the new parameters $\lambda$ and $\Sigma$. By comparing the hydrostatic equilibrium from the GR with Eq. (\ref{eq23}), we observe that the parameter $ \lambda $ cannot assume values in the interval $1/2<\lambda<2/3$. This is related to the fact that the star mass becomes negative in this range. In the next section we use Eq. (\ref{eq23}) to study the gravitational equilibrium of neutron stars. 

\section{Results and discussion}
\label{results}

We next analyze the effects of the modifications of Rastall-Rainbow approach on neutron star properties in order to test the new theory. We first analyze separately the effects of each component of the new theory,  i.e., for each EoS used in this work, we vary just the $\Sigma$ parameter while keeping $\lambda$ fixed and then 
we vary $\lambda$ while keeping $\Sigma$ fixed. 
After this test, both parameters are allowed to vary around the ranges proposed in \cite{fabris_rastall} and \cite{hendi}, i.e., $ \lambda $ cannot assume values in the interval $1/2<\lambda<2/3$ and $\Sigma$ has to be larger than 1 because lower values were shown to decrease the maximum stellar mass. We have also analyzed values of $\lambda$ smaller than 1, although they were not considered in \cite{fabris_rastall}. We have checked that for $\lambda$ values lower than certain values (always close to one), $\tilde p$ may become negative, which means that the system is unstable. If negative pressures appear only at very low densities, typical of the ones present in the inner crust, this part of the EoS can simply be eliminated. However, if the negative pressure appears at densities of the order of the ones present in the core of the EoS, the generated EoS has to be discarded. One example is given and discussed next. 

In all presented Figures the continuum line corresponds to the general relativity TOV solution and the maximum mass and radius of GR solution for each EoS can also be seen in the third column of all Tables. In the present paper we have used the full BPS EoS to describe the outer crust \cite{bps} and no hyperons are included in the EoS used next. Note that we recover GR solution in Rastall-Rainbow theory using $\Sigma=1.0$ and $\lambda =1.0$. 

\begin{figure*}[t]
\centering
\begin{tabular}{ll}
\includegraphics[width=5.5cm,angle=270]{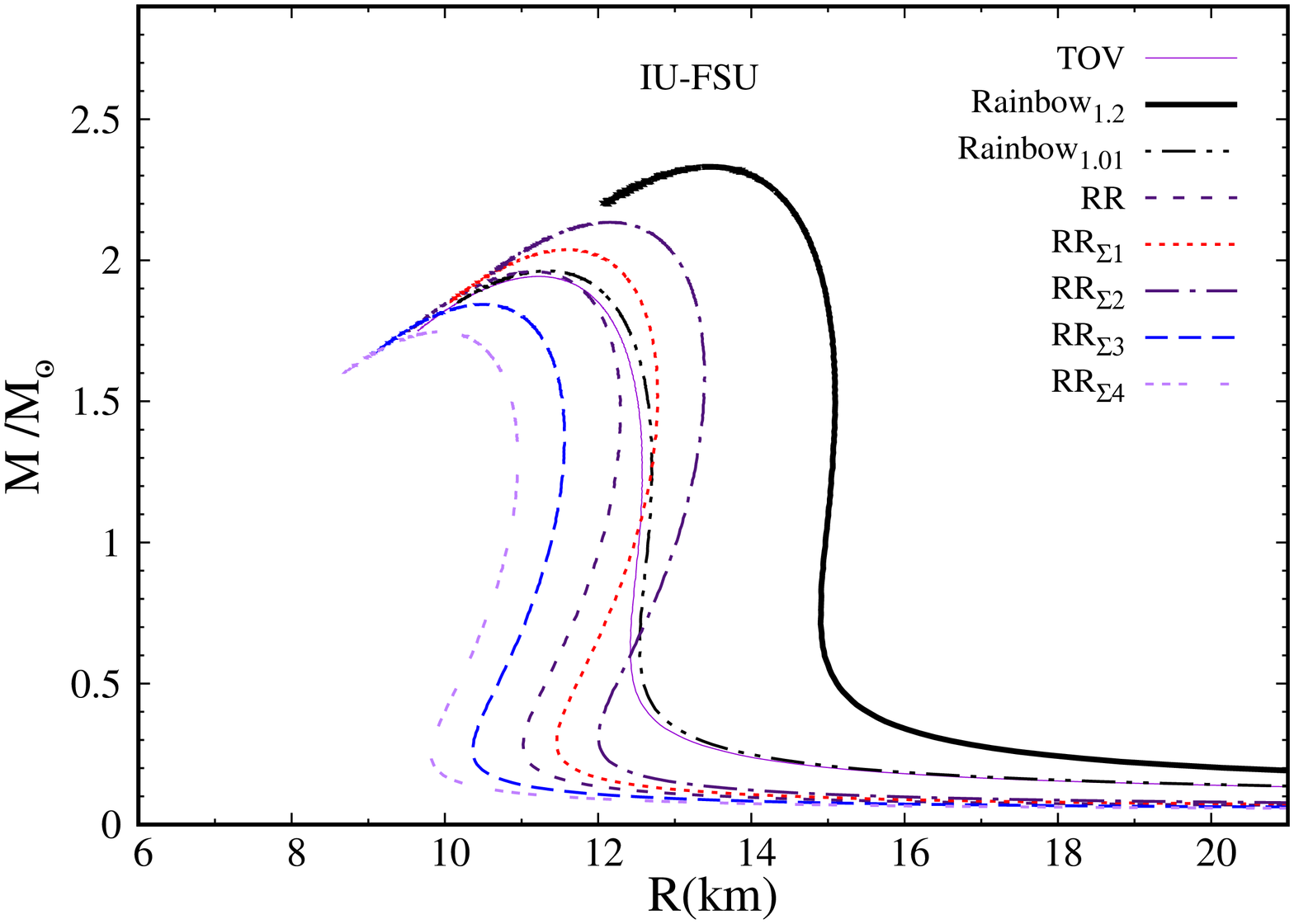}
\includegraphics[width=5.5cm,angle=270]{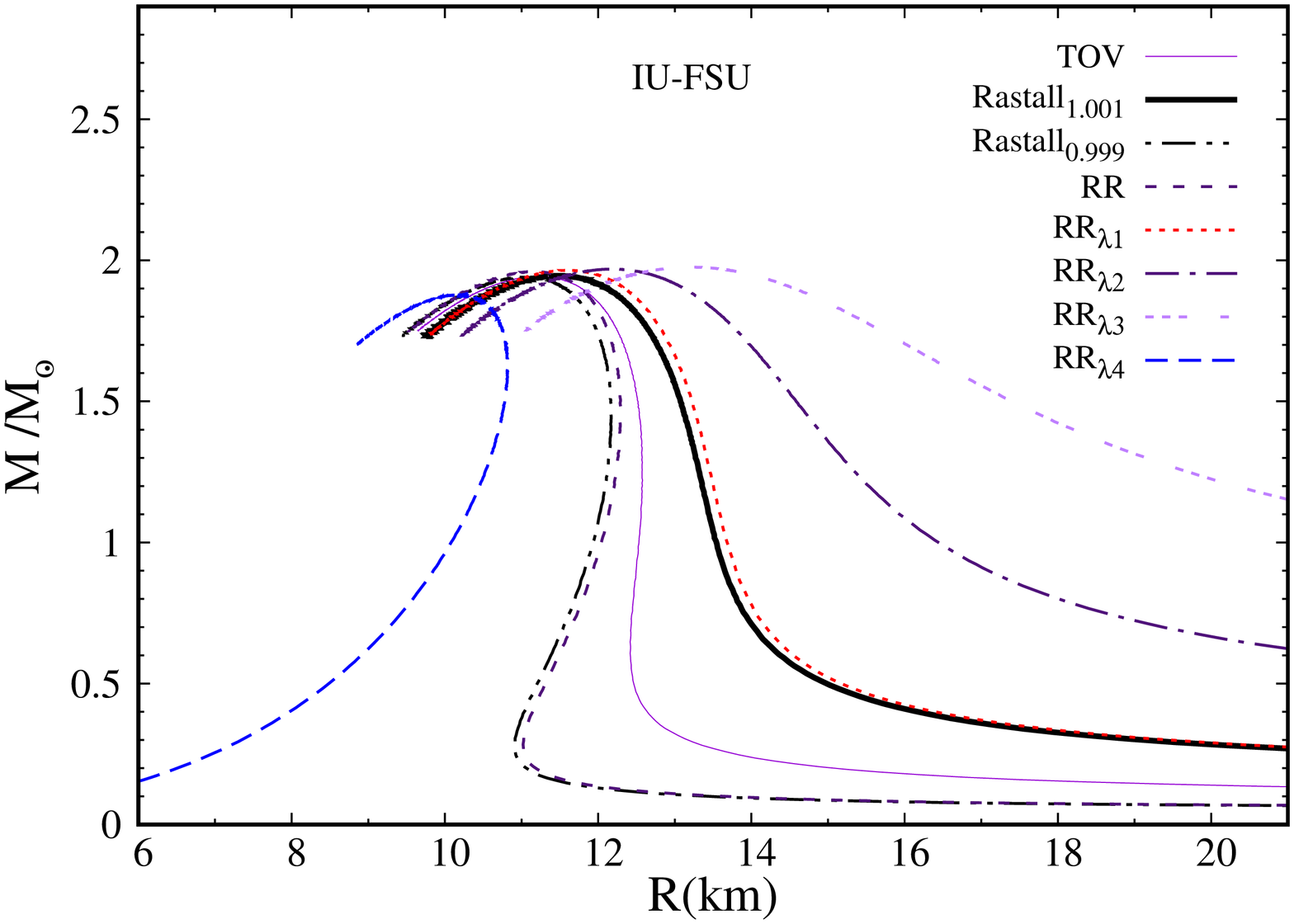} 
\end{tabular}
\caption{Mass-radius relation for a family of hadronic stars described with the IU-FSU EoS. We analyze the effects caused by varying the Rainbow parameter $\Sigma$ (left) while keeping the other parameter fixed
and the effects of varying the Rastall parameter $\lambda$ (right) while keeping $\Sigma$  fixed. 
}
\label{fig_IUFSU}
\end{figure*}

We start by testing the RR theory with one RMF model, the IU-FSU parametrization proposed in \cite{iufsu}. Besides the tests performed in \cite{Dutra2016}, IU-FSU is also successful in explaining the recent constraint that comes from the
GW170817 observation \cite{crmf}. The results obtained with various parameter values are displayed in Table \ref{tab_IUFSU} and Figure \ref{fig_IUFSU}. 
Although the results mentioned in the references above and reproduced in the present paper lie within acceptable ranges, the TOV solution of the IU-FSU EoS yields a maximum mass slightly smaller than the expected 2.0 solar masses. The radius of the canonical star (1.4M$_{\odot}$), however, lies inside the range imposed by the GW170817 constraints, which suggest that $R_{1.4}$ should lie between 10.5 and 13.4 km. Notice from Table \ref{tab_IUFSU} that, while the Rastall theory hardly affects the the maximum stellar mass, it increases the corresponding radius \cite{fabris_rastall}. The Rainbow theory, on the other hand, works in such a way that the maximum mass can either increase or decrease, depending on the values chosen. However, if the maximum mass increases, so does the radius. If it decreases, the radius also decreases \cite{hendi}. It is the combination of both approaches that allows the maximum mass to increase at the same time that the canonical star radius decreases. This feature puts the macroscopic properties obtained with the IU-FSU model comfortably within the accepted constraints for a variety of parameters.

\begin{figure*}[t]
\centering
\includegraphics[width=5.8cm,angle=270]{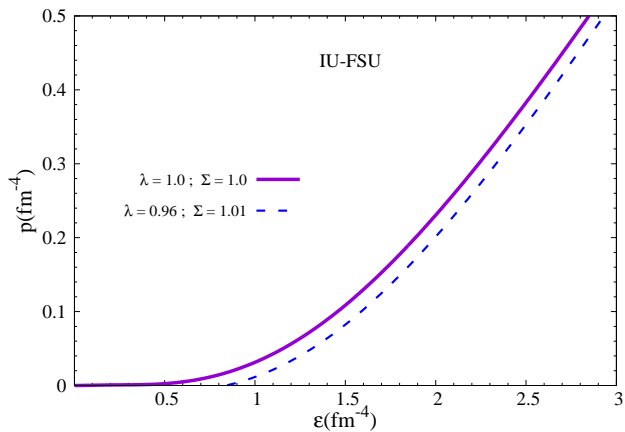}
\caption{IU-FSU EoS (solid line) and corresponding RR EoS (dashed line) obtained with the parameters named  RR$_{\lambda4}$ in Table \ref{tab_IUFSU}.}
\label{EOS_quark}
\end{figure*}

Before we investigate other EoS, we would like to comment on the result obtained with the parameter RR$_{\lambda4}$ and, for this purpose, the modified EoS is shown in Figure \ref{EOS_quark} alongside the original IU-FSU EoS. The RR TOV-like equation Eq. (\ref{eq23}) depends on modified expressions for the pressure and energy density, as given in Eq. (\ref{eq18}) and shown by the dashed curve. We can observe that the EoS is shifted towards higher densities, with a behavior equivalent to what is usually obtained with quark matter EoS without a crust \cite{quarks}. As a consequence, the resulting mass-radius diagram, depicted in Figure \ref{fig_IUFSU} right presents the typical shape of a quark star curve. All EoS with this behavior will be eliminated from our analyses and this restriction is related to the fact that $\lambda$ has a lower limit in RR theory. This lower limit was identified in \cite{fabris_rastall} as being lower than 1.0, but we have verified that the exact value is actually model dependent. 

We now investigate how model dependent the overall results are and for this purpose, other EoS are used.
The QMC \cite{guichon,saito94,saito95,pal95} and QMC$\omega \rho$ \cite{Panda12,Grams18} EoS are obtained with an effective model in which the hadrons are made of three quarks confined in a system of non-overlapping MIT bags. In the QMC models the quarks inside the nucleons interact with each other
trough the exchange of $\sigma$, $\rho$ and $\omega$ mesons. The difference between the two models comes from the fact that only in the QMC$\omega \rho$ the mesons $\omega$ and $\rho$ interact with each other, while in the standard QMC model all mesons interact just with the quarks. 
This interaction has the effect of decreasing the slope of the symmetry energy of the model
and as a consequence shrink the canonical 1.4 M$_{\odot}$ radius. More details of the QMC and QMC$\omega \rho$ models and their effects on neutron star properties can be seen in \cite{Grams18}.

The TOV solutions for the QMC and QMC$\omega \rho$ models give maximum masses within the acceptable range of $2.0 \leq M_{max} \leq 2.3$ \cite{Demorest,Antoniadis,Margalit,Shibata,Rezzolla}. Also, both EoS produce canonical $1.4$M$_{\odot}$ radii within the recent range obtained in \cite{malik} of $11.82$ km $ \leq R_{1.4 M_{\odot}} \leq 13.72$ km but slightly out of the range proposed by \cite{GW170817} of  $10.5$ km $ \leq R_{1.4 M_{\odot}} \leq 13.4$ km. 

\begin{figure*}[t]
\centering
\begin{tabular}{ll}
\includegraphics[width=5.5cm,angle=270]{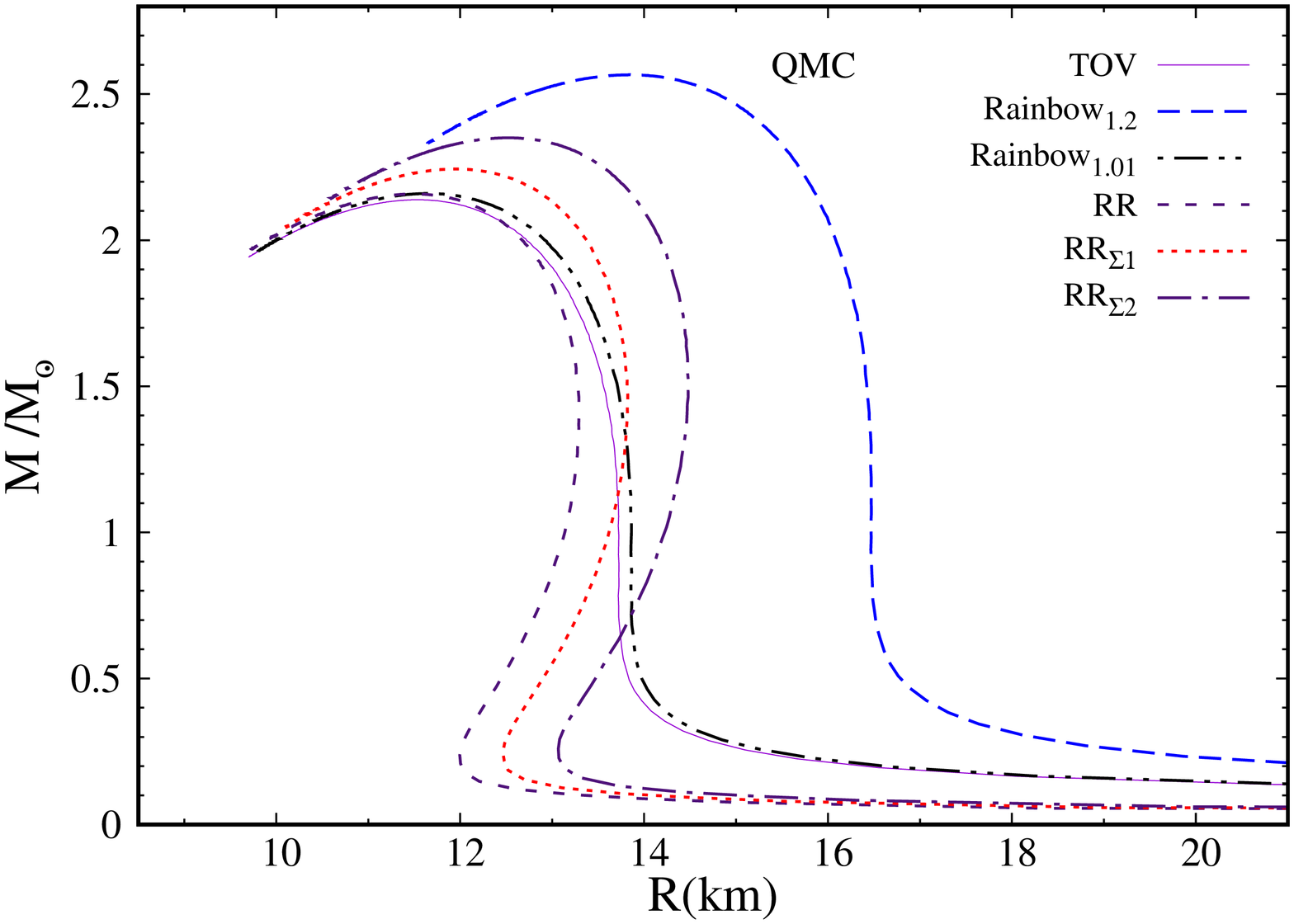}
\includegraphics[width=5.5cm,angle=270]{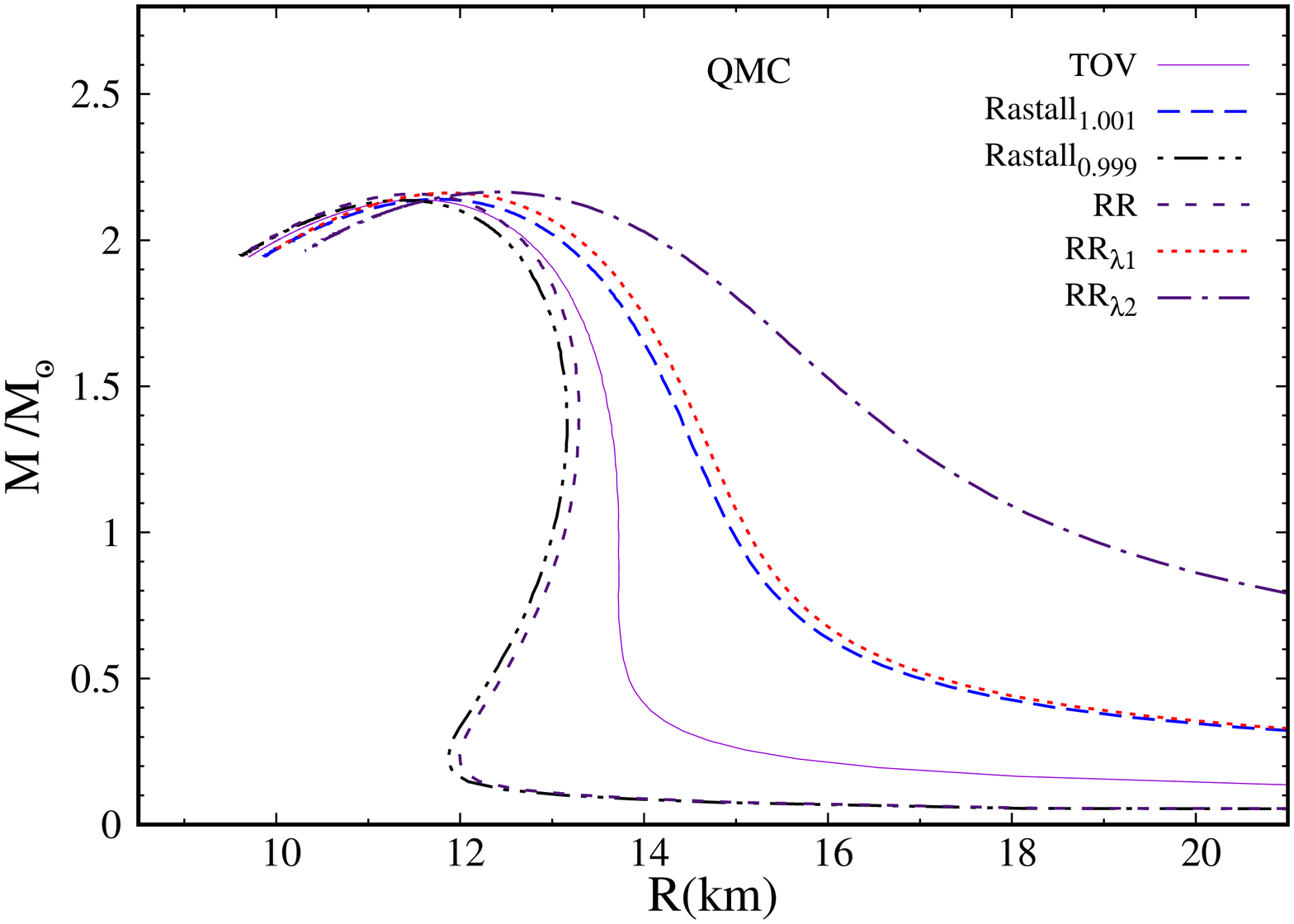} 
\end{tabular}
\caption{Mass-radius relation for a family of hadronic stars described with the QMC EoS. We analyze the effects caused by varying the Rainbow parameter $\Sigma$ (left) while keeping the other parameter fixed
and the effects of varying the Rastall parameter $\lambda$ (right) while keeping $\Sigma$  fixed. 
}
\label{fig1}
\end{figure*}

In Figure \ref{fig1} and Table \ref{tab1} we show the effects of RR theory in neutron star properties obtained with the QMC EoS. We have followed the same procedure as for the IU-FSU model, i.e., we have tested the $\Sigma$ parameter from 0.7 to 1.4 and $\lambda$ from 0.9 to 1.1, but just the results close to the accepted observational constraints are shown. For 
$\lambda = 1.006$, for instance, we have already a radius of 20.84 km and any value bigger than this provides unreasonable results when applied to neutron stars.
For $\lambda > 1$ the very small corrections on GR show no effect on the maximum stellar mass, but produce a big effect on the canonical $1.4$M$_{\odot}$ neutron star radius.  This result is in agreement with \cite{fabris_rastall} that concluded that, when confronted with neutron star constraints, just small corrections of GR coming from Rastall theory are allowed. 
Again, we have seen that for certain combinations, the maximum mass increases and the stellar radius decrease.

\begin{figure*}[t]
\centering
\begin{tabular}{ll}
\includegraphics[width=5.5cm,angle=270]{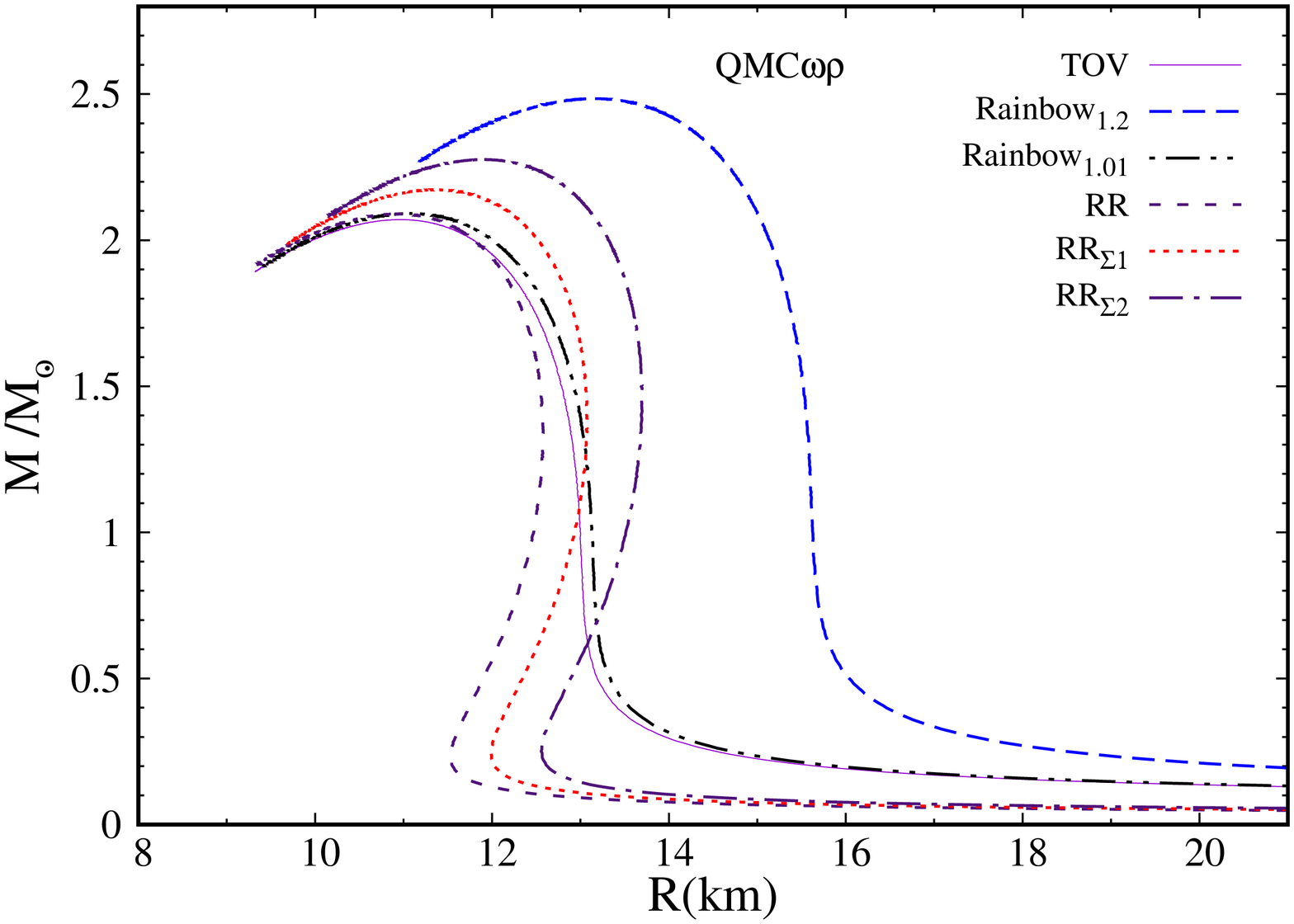}
\includegraphics[width=5.5cm,angle=270]{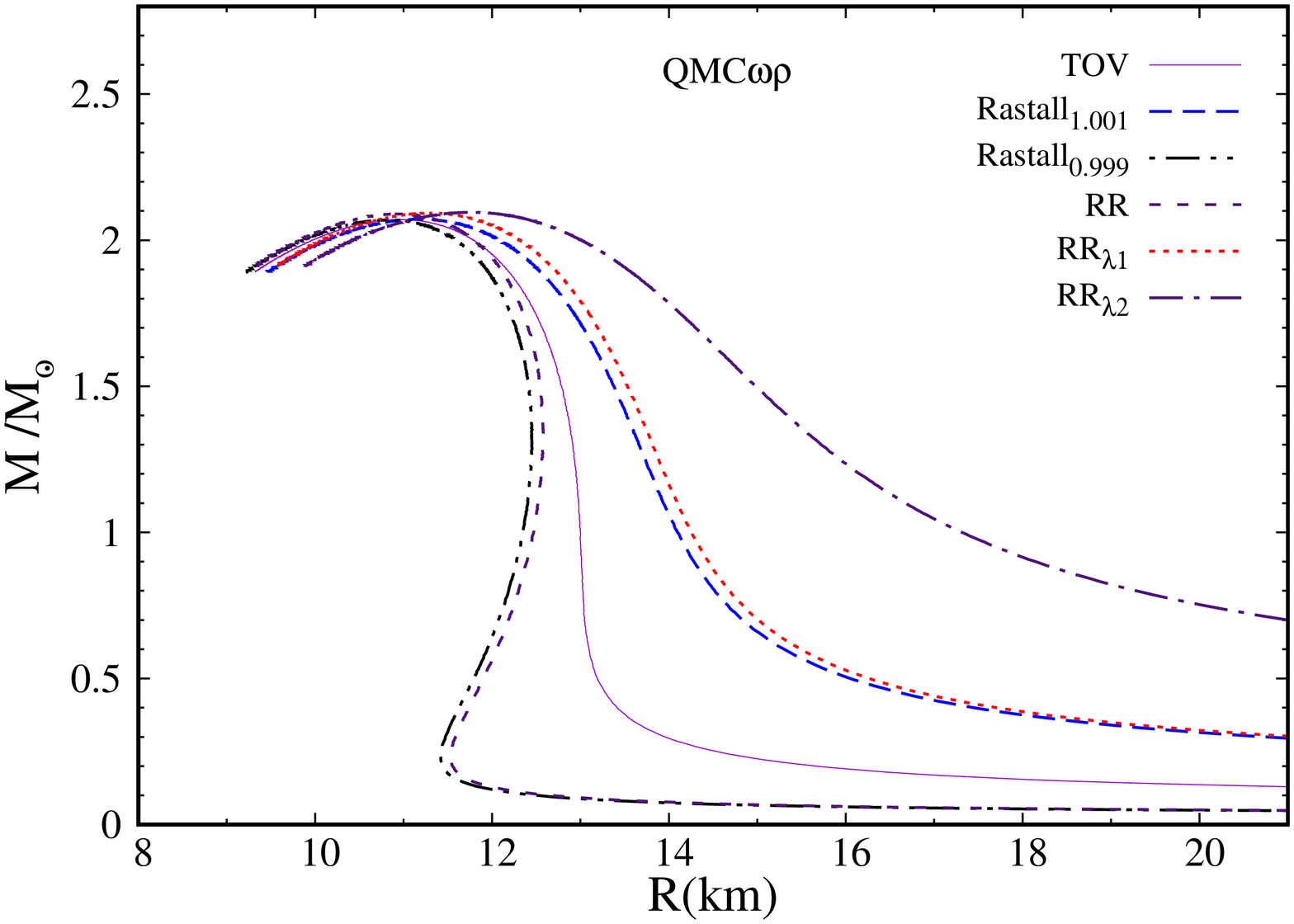} 
\end{tabular}
\caption{Mass-radius relation for a family of hadronic stars described with the QMC$\omega \rho$ EoS. We analyze the effects caused by varying the Rainbow parameter $\Sigma$ (left) while keeping the other parameter fixed
and the effects of varying the Rastall parameter $\lambda$ (right) while keeping $\Sigma$  fixed. 
}
\label{fig2}
\end{figure*}

As a final check, the QMC$\omega \rho$ model is also used to test the RR theory and the results are shown in Table \ref{tab2} and Figure \ref{fig2}. The conclusions are again the same as the ones discussed above for the IU-FSU and QMC models.

\section{Final Remarks}
\label{conclusion}

In this paper, we have combined the Rastall and Rainbow theories of modified gravity and shown that the effect of Rainbow gravity can be incorporated into the Rastall field equations by considering an energy dependent metric and an energy dependent gravitational constant.  We consider a line element with spherical symmetry and assume that the matter in the stellar interior can be described by the tensor energy-moment of a perfect fluid. In this work, we have also derived the TOV-analogue for a compact star described by hydrostatic equilibrium equations within the proposed Rastall-Rainbow gravity. 

Three RMF models have been used to test the new theory, namely IU-FSU, QMC and QMC$\omega \rho$, all of them already presenting macroscopic properties not too far away from the currently expected values. While IU-FSU reproduces well the canonical star radius, it does not reach the maximum stellar mass. On the other hand, QMC and QMC$\omega \rho$ reach the 2$M_\odot$ mass value, but the radii are a bit too large.

We have checked that while the Rastall theory alone affects very little the maximum stellar mass, it increases the corresponding radius, as already pointed out in \cite{fabris_rastall}. We have also confirmed that to avoid instabilities in the pressure, only values of $\lambda$ within $0.1 \%$ of difference from General Relativity are accepted \cite{fabris_rastall}. Nevertheless values smaller than 1 for the $\lambda$ parameter are possible, but the exact number is model dependent. 

The Rainbow theory alone works in such a way that the maximum stellar mass can either increase or decrease, depending on the $\Sigma$ values chosen. However, if the maximum mass increases, so does the radius. If it decreases, the radius also presents smaller values \cite{hendi}. 

We have then verified that, independently of the model considered, it is the combination of both theories that allows the maximum mass to increase at the same time that the canonical star radius decreases. Within this new framework, all models studied can produce macroscopic properties within the currently accepted range for a variety of parameters.

Finally, an important consideration on the hyperon puzzle should be made. In all EoS discussed in the present work, only nucleons (and leptons to insure charge neutrality and $\beta$-equilibrium) were considered. However, in very dense matter as the one existing in the interior of neutron stars, hyperons are indeed expected to appear, but their inclusion are known to soften the EoS and hence, produce lower maximum masses. One possible way to reconcile the recent measurements of massive stars with relatively small radii, is to incorporate either strange mesons or a new degree of freedom (not necessarily  known) in the calculations \cite{Luiz2018}. Had we included hyperons in our calculations with all three models investigated in the present work, they would all fail to describe 2$M_\odot$ stars. A clear way of circumventing this puzzle is the use of the RR theory we propose in the present work.

%%%%%%%%%%%%%%%%%%%%%%%%%%%%%%%%%%%%%%%%%%%%%%%%%%%%
%\begin{landscape}
\begin{table*}[t]
\centering
\caption{Macroscopic properties for different values of the $\lambda$ and $\Sigma$ parameters corresponding to the mass-radius diagram in FIG.\ref{fig_IUFSU}.}
{\small
\begin{tabular}{c|c|c|ccccccc}
\hline
\hline
\textbf{Rainbow} & Model & \ \ TOV \ \ & Rainbow$_{1.2}$ \ \cite{hendi} \ \ & Rainbow$_{1.01}$ \ \ & RR \ \ & RR$_{\Sigma1}$ \ \ & RR$_{\Sigma2}$ \ \ & RR$_{\Sigma3}$ \ \ & RR$_{\Sigma4}$ \tabularnewline 

\hline
 & $\mathbf{\Sigma}$  &  \textbf{1.0}  & \textbf{1.2}  & \textbf{1.01} & \textbf{1.01}  &  \textbf{1.05}  & \textbf{1.1} & \textbf{0.95}  & \textbf{0.90}  \tabularnewline
 
 Parameters & $\lambda$  &  1.0  &  1.0 &  1.0 &  0.999  & 0.999  &  0.999 & 0.999 & 0.999  \tabularnewline
 
\hline
& $M_{max}$  &  1.94 $M_\odot$ &  2.33 $M_\odot$  & 1.96 $M_\odot$  &  1.96 $M_\odot$  & 2.03 $M_\odot$ & 2.13 $M_\odot$ & 1.84 $M_\odot$ & 1.74 $M_\odot$\tabularnewline

IU-FSU & $R_{M_{max}}$ &  11.22 km  \ \ &  13.46 km   \ \ & 11.33 km \ \ & 11.15 km  \ \ & 11.59 km  \ \  & 12.15 km  \ \ & 10.49 km \ \ & 9.94 km \ \ \tabularnewline

 &$R_{1.4}$& 12.55 km  \ \ &  15.08 km \ \ & 12.68 km  \ \ & 12.28 km  \ \ & 12.76 km  \ \ & 13.34 km  \ \ & 11.55 km \ \ & 10.92 km \ \ \tabularnewline
\hline
 &  &   \ \ &    \ \ &\ \ &   \ \ &    \ \  &   \ \ &  \ \ \tabularnewline

 &  &  General  \ \ &    \ \ &\ \ &   \ \ & Modified   \ \  &   \ \ &  \ \ \tabularnewline
  &  &  Relativity  \ \ &    \ \ &\ \ &   \ \ & Gravity \ \  &   \ \ &  \ \ \tabularnewline
   &  &    \ \ &     \ \ &\ \ &   \ \ &    \ \  &   \ \ &  \ \ \tabularnewline

\hline
\textbf{Rastall} & Model & \ \ TOV \ \ \ & Rastall$_{1.001}$ \ \cite{fabris_rastall} \ \ \ & Rastall$_{0.999}$ \ \ \ & RR \ \ \ & RR$_{\lambda1}$ \ \ \ & RR$_{\lambda2}$ \ \ \ & RR$_{\lambda3}$ \ \ \ & RR$_{\lambda4}$ \tabularnewline 
\hline
& $\boldsymbol{\lambda}$  &  \textbf{1.0}  &  \textbf{1.001} &  \textbf{0.999}  &  \textbf{0.999}  & \textbf{1.001} & \textbf{1.003} & \textbf{1.006}  & \textbf{0.96} \tabularnewline
 
Parameters & $\Sigma$  &  1.0  & 1.0 & 1.0 & 1.01  & 1.01 &  1.01  & 1.01 & 1.01  \tabularnewline
\hline
& $M_{max}$  &  1.94 $M_\odot$ &  1.94 $M_\odot$  & 1.94 $M_\odot$  &  1.96 $M_\odot$  &  1.96 $M_\odot$ & 1.96 $M_\odot$  & 1.97 $M_\odot$ & 1.87 $M_\odot$\tabularnewline

IU-FSU & $R_{M_{max}}$ &  11.22 km  \ \ &  11.48 km   \ \ & 11.05 km \ \ & 11.15 km  \ \ & 11.60 km  \ \  & 12.18 km  \ \ & 13.21 km \ \ & 10.17 km \ \ \tabularnewline

 &$R_{1.4}$& 12.55 km  \ \ &  13.18 km \ \ & 12.16 km  \ \ & 12.28 km  \ \ & 13.32 km  \ \ & 14.87 km  \ \ & 18.19 km \ \ & 10.70 km \ \ \tabularnewline
\hline
\hline
\label{tab_IUFSU}
\end{tabular}
}
\end{table*}
%\end{landscape}

%%%%%%%%%%%%%%%%%%%%%%%%%%%%%%%%%%%%%%%%%%%%%%%%%%%%
%%%%%%%%%%%%%%%%%%%%%%%%%%%%%%%%%%%%%%

%%%%%%%%%%%%%%%%%%%%%%%%%%%%%%%%%%%%%%%%%%%%%%%%%%%%
\begin{table*}[t]
\centering
\caption{Macroscopic properties for different values of the $\lambda$ and $\Sigma$ parameters corresponding to the mass-radius diagram in FIG.\ref{fig1}.}
{\footnotesize
\begin{tabular}{c|c|c|ccccccc}
\hline
\hline
\textbf{Rainbow} & Model & \ \ TOV \ \ & Rainbow$_{1.2}$ \ \ \cite{hendi} \ \ & Rainbow$_{1.01}$ \ \ & RR \ \ \ & RR$_{\Sigma1}$ \ \ \ & RR$_{\Sigma2}$ \tabularnewline 
\hline

 & $\mathbf{\Sigma}$  &  \textbf{1.0}  & \textbf{1.2}  & \textbf{1.01} & \textbf{1.01}  &  \textbf{1.05}  & \textbf{1.1}  \tabularnewline
 
Parameters & $\lambda$  &  1.0  &  1.0 & 1.0 & 0.999  &  0.999  & 0.999  \tabularnewline
 
\hline
& $M_{max}$  &  2.14 $M_\odot$ &  2.56 $M_\odot$ &  2.15 $M_\odot$  & 2.15 $M_\odot$  &  2.24 $M_\odot$  & 2.35 $M_\odot$  \tabularnewline

QMC & $R_{M_{max}}$ &  11.53 km  \ \ &  13.85 km  \ \ & 11.65 km \ \ & 11.49 km \ \ & 11.95 km  \ \ & 12.51 km  \ \  \tabularnewline

&$R_{1.4}$& 13.61 km  \ \ &  16.44 km \ \ & 13.76 km \ \ & 13.28 km  \ \ & 13.80 km  \ \ & 14.46 km  \ \ \tabularnewline
\hline
 &  &          \ \ &    \ \ &\ \ &   \ \ &    \ \  &   \ \ &  \ \ \tabularnewline

 &  &  General   \ \ &    \ \ &\ \ & Modified   \ \ &    \ \  &   \ \ &  \ \ \tabularnewline
  &  &  Relativity  \ \ &    \ \ &\ \ & Gravity \ \ &    \ \  &   \ \ &  \ \ \tabularnewline
   &  &            \ \ &     \ \ &\ \ &   \ \ &    \ \  &   \ \ &  \ \ \tabularnewline

\hline
\textbf{Rastall} & Model & \ \ TOV \ \ \ & Rastall$_{1.001}$ \ \cite{fabris_rastall} \ \ \ & Rastall$_{0.999}$ \ \ \ & RR \ \ \ & RR$_{\lambda1}$ \ \ \ & RR$_{\lambda2}$ \tabularnewline 

\hline
 & $\boldsymbol{\lambda}$  &  \textbf{1.0}  &  \textbf{1.001}  &  \textbf{0.999} &  \textbf{0.999}  & \textbf{1.001} & \textbf{1.003} \tabularnewline
 
Parameters & $\Sigma$  &  1.0  & 1.0  & 1.0 & 1.01  &  1.01  & 1.01 \tabularnewline
 
\hline
& $M_{max}$  &  2.14 $M_\odot$ &  2.14 $M_\odot$ &  2.13 $M_\odot$ & 2.15 $M_\odot$  &  2.16 $M_\odot$  & 2.16 $M_\odot$  \tabularnewline

QMC & $R_{M_{max}}$ &  11.53 km  \ \ &  11.77 km \ \ & 11.37 km  \ \ & 11.49 km \ \ & 11.88 km  \ \ & 12.39 km  \ \ \tabularnewline

 &$R_{1.4}$& 13.61 km  \ \ &  14.39 km \ \ &  13.16 km \ \ & 13.28 km  \ \ & 14.54 km  \ \ & 16.48km  \ \ \tabularnewline
\hline
\hline
\end{tabular}
}
\label{tab1}
\end{table*}

%%%%%%%%%%%%%%%%%%%%%%%%%%%%%%%%%%%%%%%%%%%%%%%%%%%%%%%%%%%%%%%%%%%%%%%%%%%%%%%%%%%%%%%%%%%%

%%%%%%%%%%%%%%%%%%%%%%%%%%%%%%%%%%%%%%%%%%%%%%%%%%%%%%%%%%%%%%%%%%%%%%%%%%%%%%%%%%%%%%%%%%%%
%%%%%%%%%%%%%%%%%%%%%%%%%%%%%%%%%%%%%%%%%%%%%%%%%%%%%%%%%%%%%%%%%%%%%%%%%%%%%%%%%%%%%%%%%%%%

\begin{table*}[t]
\centering
\caption{Macroscopic properties for different values of the $\lambda$ and $\Sigma$ parameters corresponding to the mass-radius diagram in FIG.\ref{fig2}.}
{\footnotesize
\begin{tabular}{c|c|c|cccccc}
\hline
\hline
\textbf{Rainbow} & Model & \ \ TOV \ \ \ & Rainbow$_{1.2}$ \ \cite{hendi} \ \ \ & Rainbow$_{1.01}$ \ \ \ & RR \ \ \ & RR$_{\Sigma1}$ \ \ \ & RR$_{\Sigma2}$ \tabularnewline 
\hline
 &  $\mathbf{\Sigma}$  &  \textbf{1.0}  & \textbf{1.2}  & \textbf{1.01} & \textbf{1.01}  &  \textbf{1.05}  & \textbf{1.1} \tabularnewline
 
Parameters & $\lambda$  &  1.0  & 1.0 & 1.0 &  0.999  & 0.999  &  0.999 \tabularnewline
 
\hline
& $M_{max}$  &  2.07 $M_\odot$ &  2.48 $M_\odot$ & 2.09 $M_\odot$ & 2.09 $M_\odot$  &  2.17 $M_\odot$  & 2.27 $M_\odot$ \tabularnewline

QMC$\omega \rho$ & $R_{M_{max}}$ &  10.96 km  \ \ &  13.15 km \ \ &  11.07 km  \ \ & 10.93 km \ \ & 11.36 km  \ \ & 11.90 km  \ \ \tabularnewline

 &$R_{1.4}$& 12.83 km  \ \ &  15.55 km \ \ &  12.99 km \ \ & 12.56 km  \ \ & 13.07 km  \ \ & 13.68 km  \ \ \tabularnewline
\hline
 &  &          \ \ &    \ \ &\ \ &   \ \ &    \ \  &   \ \ &  \ \ \tabularnewline

 &  &  General   \ \ &    \ \ &\ \ & Modified  \ \ &   \ \  &   \ \ &  \ \ \tabularnewline
  &  &  Relativity  \ \ &    \ \ &\ \ & Gravity  \ \ &  \ \  &   \ \ &  \ \ \tabularnewline
   &  &            \ \ &     \ \ &\ \ &   \ \ &    \ \  &   \ \ &  \ \ \tabularnewline

\hline
\textbf{Rastall} & Model & \ \ TOV \ \ \ & Rastall$_{1.001}$ \ \cite{fabris_rastall} \ \ \ & Rastall$_{0.999}$ \ \ \ & RR \ \ \ & RR$_{\lambda1}$ \ \ \ & RR$_{\lambda2}$ \tabularnewline 
\hline
& $\boldsymbol{\lambda}$  &  \textbf{1.0}  &  \textbf{1.001} &  \textbf{0.999} &  \textbf{0.999}  & \textbf{1.001} & \textbf{1.003} \tabularnewline
 
Parameters & $\Sigma$  &  1.0  & 1.0 & 1.0 & 1.01  &  1.01  & 1.01 \tabularnewline
 
\hline
& $M_{max}$  &  2.07 $M_\odot$ &  2.07 $M_\odot$ &  2.06 $M_\odot$ & 2.09 $M_\odot$  &  2.09 $M_\odot$  & 2.09 $M_\odot$ \tabularnewline

QMC$\omega \rho$ & $R_{M_{max}}$ &  10.96 km  \ \ &  11.18 km \ \  &  10.82 km  \ \ & 10.93 km \ \ & 11.29 km  \ \ & 11.76 km  \ \  \tabularnewline

 &$R_{1.4}$& 12.83 km  \ \ &  13.52 km \ \ &  12.43 km \ \ & 12.56 km  \ \ & 13.67 km  \ \ & 15.32 km  \ \ \tabularnewline
\hline
\hline
\end{tabular}
}
\label{tab2}
\end{table*}
%%%%%%%%%%%%%%%%%%%%%%%%%%%%%%%%%%%%%%%%%%%%%%%%%%%%%%%%%%%%%%%%%%%%%%%%%%%%%%%%%%%%%%%%%%%%

%%%%%%%%%%%%%%%%%%%%%%%%%%%%%%%%%%%%%%%%%%%%%%%%%%%%%%%%%%%%%%%%%%%%%%%%%%%%%%%%%%%%%%%%%%%%
%%%%%%%%%%%%%%%%%%%%%%%%%%%%%%%%%%%%%%%%%%%%%%%%%%%%%%%%%%%%%%%%%%%%%%%%%%%%%%%%%%%%%%%%%%%%
%%%%%%%%%%%%%%%%%%%%%%%%%%%%%%%%%%%%%%%%%%%%%%%%%%%%%%%%%%%%%%%%%%%%%%%%%%%%%%%%%%%%%%%%%%%%

\acknowledgments
\noindent This work is a part of the project CNPq-INCT-FNA Proc. No. 464898/2014-5. DPM acknowledges partial support from CNPq (Brazil) under grant 301155/2017-8 and CEM has a scholarship paid by Capes (Brazil).

\end{document}